\RequirePackage{fix-cm}
\documentclass[smallextended]{svjour3}       
\smartqed  

\usepackage{graphicx}
\usepackage{epstopdf}
\usepackage{lineno,hyperref}
\usepackage{amsmath}
\usepackage{caption}
\usepackage{cite}
\usepackage{graphicx}
\usepackage{subfigure}
\usepackage{verbatim}
\usepackage{epstopdf}
\usepackage{url}

\usepackage{geometry}
\begin{document}

\title{Security analysis of a self-embedding fragile image watermark scheme
}


\author{Xinhui Gong         \and
        Feng Yu             \and
        Xiaohong Zhao       \and
        Shihong Wang
}




\institute{Shihong Wang \at
              School of Science, Beijing University of Posts and Telecommunications, Beijing 100876, China \\
              Tel.: 0086-10-62282452\\
              Fax: 0086-10-62282452\\
              \email{shwang@bupt.edu.cn}           
           \and
}

\date{Received: date / Accepted: date}

\maketitle

\begin{abstract}
Recently, a self-embedding fragile watermark scheme based on reference-bits interleaving and
adaptive selection of embedding mode was proposed. Reference bits are derived from the scrambled MSB bits of a cover image, and then are combined with authentication bits to form the watermark bits for LSB embedding. We find this algorithm has a feature of block independence of embedding watermark such that it is vulnerable to a collage attack. In addition, because the generation of authentication bits via hash function operations is not related to secret keys, we analyze this algorithm by a multiple stego-image attack. We find that the cost of obtaining all the permutation relations of $l\cdot b^2$ watermark bits of each block (i.e., equivalent permutation keys) is about $(l\cdot b^2)!$ for the embedding mode $(m, l)$, where $m$ MSB layers of a cover image are used for generating reference bits and $l$ LSB layers for embedding watermark, and $b\times b$ is the size of image block. The simulation results and the statistical results demonstrate our analysis is effective.
\end{abstract}

\section{Introduction}
\label{Sec. 1}
With the development in science and technology, digital images are easily processed and widely used \cite{Birajdar2013Digital, Petitcolas1999Information, Li2011A, Ping2016Chaos}. To ensure the integrity of digital images, image authentication techniques based on fragile watermark have been studied \cite{Lin2005A,Ping1998A}. Haouzia et al. \cite{Haouzia2008Methods} discussed the general requirements of an authentication system, such as its security, the sensitivity of watermark, the accuracy of tampering localization. We consider security the most important one. Any authentication system must protect the authentication data against any falsification attempts.

There are multiple image authentication schemes based on fragile watermark proposed by scholars\cite{Liu2007An, Lazarov2016A, Qin2017Fragile, Zhang2007Statistical, Zhang2017FragileA, Zhang2009Fast, Zhang2017Fragile}. However, some of them only pursue the accuracy of tampering detection and the quality of recovered image. There still exist the security problems, which are vulnerable to counterfeiting attack. Yeung and Mintzer  \cite{Yeung1997An} proposed a fragile watermark scheme where fragile watermark is generated via a lookup table. The table map the value of pixels to 0 or 1 bit controlled by the secret keys. However, there is a security risk that the mapping relations are not related to the image content. Holliman and Memon proposed a vector quantization (VQ) attack to break this scheme\cite{Holliman2000Counterfeiting}. Chang et al. \cite{Chang2006A} proposed a watermark algorithm based on hash functions. The authentication bits are generated via a cryptographic hash function, and then inserted into the lowest significant bit (LSB) of the center pixel in a corresponding block. However, Phan \cite{Phan2008Tampering} proposed an effective method to break this scheme. Lin et al.\cite{Lin2005A} proposed a hierarchical watermark method, where the feature of each block of the image is embedded into another block. This scheme uses both reference bits and authentication bits to detect tampered area. However, Chang et al. \cite{Chang2008Four} proposed a four-scanning attack to find the block-mapping sequence, and furthermore to counterfeit authenticated images successfully. A scheme with high data hiding capability and fidelity preservation was proposed by Lin et al. \cite{Lin2011Protecting}. However Li et al. \cite{Li2016Attack} proposed an analysis method to counterfeit authenticated images. Rawat and Raman \cite{Rawat2011A} proposed a scheme based on chaotic map and Teng et al. \cite{Teng2013Cryptanalysis} found this scheme cannot resist a content-only attack. Without knowing the secret key, an attacker first stores the LSBs of the watermarked image. Then he or she alters the pixels and replace their LSBs stored before. The main reason is that the authentication data is irrelevant to the cover image.

Qin et al. \cite{Qin2016Self} proposed a self-embedding fragile watermark scheme based on reference-data interleaving and adaptive selection of embedding mode. The choice of embedding mode is related with watermarked image quality, estimated tampering rate, and recovered image quality. The authors claimed that the proposed scheme can achieve good visual quality of recovered images under different tampering rates. To improve the sensitivity of authentication data in the scheme, hash function operations are utilized to generate authentication data. Though this scheme is flexible and has good ability in recovering tampered image, we find it vulnerable to a collage attack. The scheme has a fatal defect that the authentication bits of each block only embed into the corresponding block. This feature is called block independence. This weakness makes this scheme not resist the collage attack. Besides, the generation of authentication bits via hash function operations is not controlled by secret keys and the authentication bits embed in the fixed positions that are not related with the cover image. Therefore, based on the two characteristic above, if an attacker obtains embedding positions of watermark, he or she can forge any authenticated images.

The remaining parts of the paper are organized as follows. Section 2 describes the conditions of security analysis of fragile watermark. In Section 3, we introduce Qin et al.'s scheme briefly. In Section 4 we analyze the security of Qin et al.'s scheme by using the collage attack and multiple stego-image attack. Conclusion of this paper is given in Section 5.

\section{Conditions of security analysis of fragile watermark}
\label{Sec. 2}
Besides VQ attack \cite{Holliman2000Counterfeiting} and collage attack \cite{Fridrich2002Cryptanalysis} etc, there also exist general tampering attacks, such as copy-paste attacks, deletion attacks, text insertion attacks etc \cite{Sreenivas2017Fragile}. Considering analysis conditions, the attack methods can be classified into the following types \cite{Fridrich2002Security}.

\textit{Stego-image attacks.} The attacker has only one authenticated image. The aim is to modify the image such that it is undetected or obtain some secret information of the scheme.

\textit{Multiple stego-image attacks.} The attacker has multiple authenticated images. The aim is to modify or forge one image such that it is undetected, or to obtain some secret information of the scheme.

\textit{Verification device attacks.} The attacker has access to the verification device, i.e., the attacker can verify the authenticity of any image. In this condition, the attacker is interested in making undetected changes or obtaining some secret information of the scheme.

\textit{Cover-image attacks.} The attacker has multiple pairs of original-authenticated images. Again, the attacker is interested in making undetected changes or obtaining some secret information of the scheme.

\textit{Chosen cover-image attacks.} The attacker has access to the authentication device and can submit her or his images for authentication. The aim is to obtain some information about the secret authentication key.

\section{Brief description of Qin et al.'s scheme}
\label{Sec. 3}
First, we describe Qin et al.'s scheme briefly. This scheme is a self-embedding image authentication scheme. In the original scheme, $m$ MSB (the most significant bit) layers of a cover image are used for the generation of reference bits, and the reference bits and the $m'$ MSB layers are further used for the generation of authentication bits ($m'$ is the minimum of two values $m$ and $8-l$, where $l$ LSB layers are used for embedding).  The authentication and reference bits are embedded in $l$ LSB layers of the cover image. There exist two working modes: overlapping-free embedding ($m+l\leq 8$) and overlapping embedding ($m+l> 8$). The choices of $m$ and $l$ are related to many factors, such as watermarked image quality, estimated tampering rate, and recovered image quality.
\subsection{Watermark embedding}
\label{Sec. 3.1}
In this scheme, the embedding and detecting of watermark are based on image block. For a block size of $b\times b$, $l\cdot b^2$ bits are embedded in $l$ LSB layers of each block, containing $L_a$ authentication bits and $l\cdot b^2-L_a$ reference bits. The generation of reference and authentication bits, and the watermark embedding procedure are shown below.

\textbf{Step 1.} Image grouping. Divide a cover image of size $N_1\times N_2$ into $N/b^2$ blocks, where $N = N_1\times N_2$. For simplicity, $N_1$ and $N_2$ are assumed to be the multiples of $b$.

\textbf{Step 2.} Permutation. Collect $m$ MSB layers of the cover image, and permute the $m\cdot N$ bits with a secret key to form a set \textbf{C}.

\textbf{Step 3.} Generate reference bits. Divide \textbf{C} into $S$ subsets noted as $\textbf{C}_1$, $\textbf{C}_2$, ... , $\textbf{C}_S$. Each subset contains $u$ bits. Each subset is transformed by the following expression
\begin{equation}
\begin{bmatrix} r_{j,1} \\ r_{j,2} \\ ...  \\ r_{j,v} \end{bmatrix}\\=
\textbf{H}_j \cdot \begin{bmatrix} c_{j,1} \\ c_{j,2} \\ ...  \\ c_{j,u} \end{bmatrix}\\
\end{equation}
where $\textbf{H}_j$ is a pseudo-random binary matrix of size $v\times u$ produced from a secret key. After the transformation above, there are $v\cdot S = v\cdot m\cdot N/u $ bits named the reference bits, furthermore, the value of $v$ should satisfy the expression
\begin{equation}
v\cdot S = v\cdot m\cdot N/u = l\cdot N-L_a\cdot N/b^2
\end{equation}

\textbf{Step 4.} Generate authentication bits. For each block, feed $m'\cdot b^2$ bits of the $m'$ MSB layers and corresponding $l\cdot b^2-L_a$ reference bits into a hash function and generate $L_a$ authentication bits.

\textbf{Step 5.} Embed watermark. Permute the $l\cdot b^2$ watermark bits of each block with a secret key (containing $L_a$ authentication bits and $l\cdot b^2-L_a$ reference bits) and use them to replace the $l$ LSB layers of each block.

\subsection{Tampering detection}
In this scheme, a receiver not only verifies the integrity of suspicious watermarked image $I_w^*$, but also has ability to recover tampering area. Here we only describe the tampering detection procedure.

For each $b\times b$ block of $I_w^*$, first extract the $l\cdot b^2$ bits of its $l$ LSB layers consisting of $L_a$ authentication bits and $l\cdot b^2-L_a$ reference bits. There are two modes of tampering detection:

(1) \textit{Overlapping-free embedding:} Feed the $m\cdot b^2$ MSB bits of each block and the $l\cdot b^2-L_a$ reference bits into a hash function and output $L_a$ authentication bits. If the recalculated $L_a$ bits differ from the extracted $L_a$, this block is judged as a tampered block. Otherwise, it is marked as an intact block.

(2) \textit{Overlapping embedding:} Feed the $(8-l)\cdot b^2$ MSB bits of each block and the $l\cdot b^2-L_a$ reference bits into a hash function and output $L_a$ authentication bits. Similar to overlapping-free embedding mode, if the recalculated $L_a$ bits differ from the extracted $L_a$, this block is judged as a tampered block. Otherwise, it is an intact one.

\section{Security analysis of Qin et al.'s scheme and simulation results}
\subsection{A collage attack}
One defect of this scheme is that the authentication bits generated by one block are embedded into the same block. This means that each block verification with unchanged secret keys is independent. This defect leads to be vulnerable to attack of a collage of images. The collage attack is effective for overlapping-free embedding mode and overlapping embedding mode. The details of the collage attack are shown as follows.

\textbf{Attack aim:} The aim of this attack is to forge a new image produced from authenticated images.

\textbf{Attack Conditions:} A forger has multiple authenticated images with the same embedding mode and the unchanged secret key, although she or he does't know the key.

\textbf{Step 1.} For an authenticated image $A$, a forger first marks the block according to the block size $b\times b$ and selects an area to be tampered. And this area must be an integral multiple of a block, named $y$.

\textbf{Step 2.} Given another authenticated image $B$, which has the same authentication mode as the image $A$. Same as $A$, the authenticated image $B$ is blocked through $b\times b$ and the area $y$ of $B$ is selected.

\textbf{Step 3.} Replace the area $y$ of $A$ with the area $y$ of $B$, and finish a forged image $A'$.

\textbf{Step 4.} Verification operation. Because the tampered area $y$ is an integral multiple of the size of a block and the verification procedure is block independent, the tampered image $A'$, as an intact image, can pass tampering detection and cannot be found that it has been tampered.

To verify our analysis above, we give three examples by using the collage attack. The simulation results are shown in Fig.1 (top, middle and bottom panels). The size of all the images is $512\times 512$ and the size of a block is $b \times b=2\times 2$. Figures 1(a1-d1) are four authenticated images with the same embedding mode $(6,2)$ , i.e., $m=6,l=2$. We take a quarter of each image and construct a collage image shown in Fig.1(e1). The collage image (e1) can successfully pass the tampering detection process, and the detection result is shown in Fig.1(f1). In Figs. 1(a2-e2), the embedding mode is $(6,3)$ and the other conditions are the same as Figs. 1(a1-e1). The collage image (e2) also passes tampering detection, and the detection result is shown in Fig.1(f2). The two examples above show that the embedding mode does not affect the collage attack.

In Fig.1, the four images of (a3-d3) are the same as those of a1-d1, but chosen parts of the four images a3-d3 for making a new image e3 are slightly different from those of a1-d1. We select a rectangular part of each image, from pixel (1,1) to pixel (251,251) in a3, from pixel (1,252) to pixel (252,512) in b3, from pixel (252,1) to pixel (512,251) in c3, and from pixel (252,252) to pixel (512,512) in d3. Since the block size is $2\times 2$, the edge blocks of the merged image e3 are combinations of pixels of two images or four images, which are inconsistent with the original blocks. Therefore it cannot pass tampering detection. The results of tampering detection (f3) validate the analysis above. Through the result of f3, we adjust the edge of the collage image, once edge blocks match the original blocks of images, the collage image can pass tampering detection.

\begin{figure}
\centering
\includegraphics[scale=0.4]{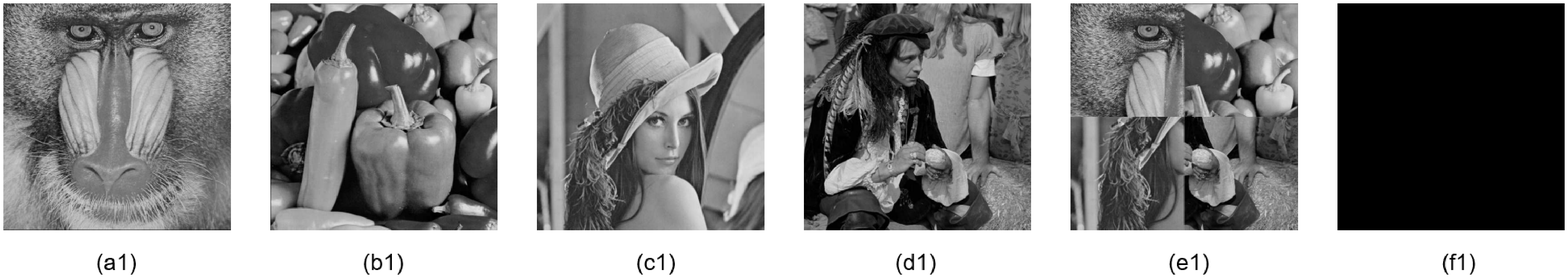}\caption*{}
\includegraphics[scale=0.4]{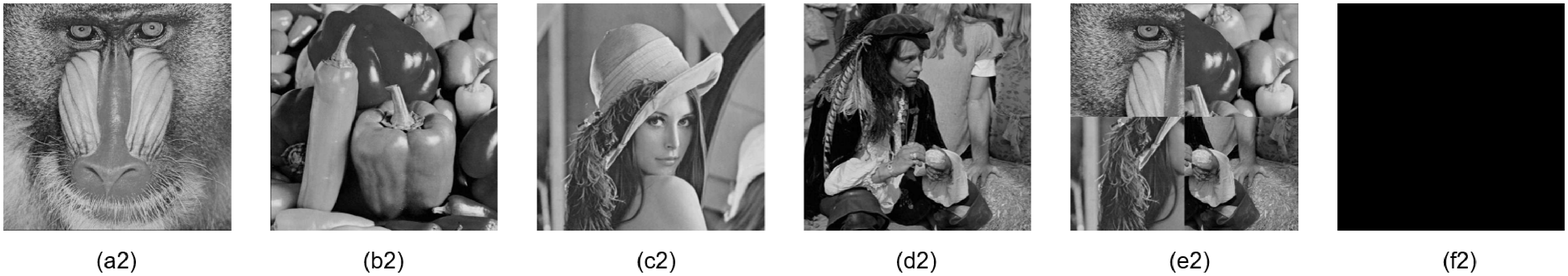}\caption*{}
\includegraphics[scale=0.4]{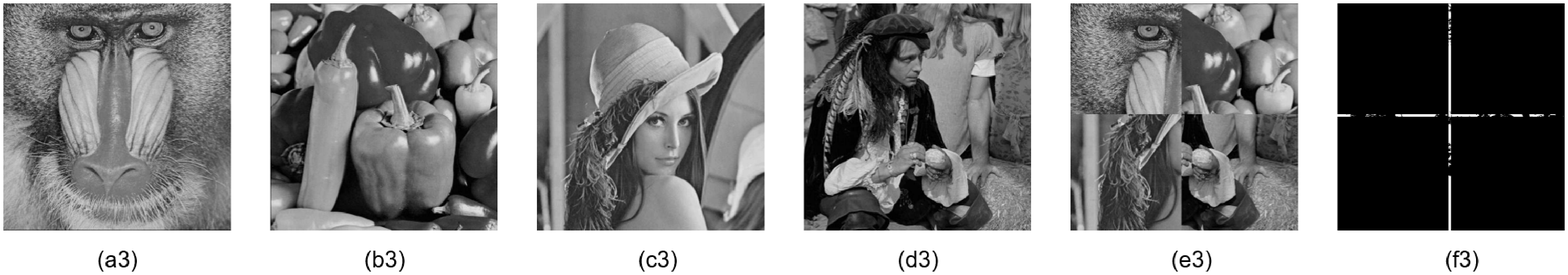}\caption*{}
  \caption{Demonstration of the collage attack for Qin et al.'s algorithm. Two examples (top and bottom panels) are embedding mode $(6, 2)$ and another one (middle panel) is embedding mode $(6, 3)$. Authenticated images (a1-d1),(a2-d2) and (a3-d3), their collage images (e1)(e2)(e3) and the tampering detection results of the collage image (f1)(f2)(f3).}
\label{fig:Fig. 1}
\end{figure}

\subsection{Multiple stego-image attack}
The aim of this attack is to obtain the equivalent permutation relation of the $l\cdot b^2$ watermark bits in each block. Once the attacker acquires these permutation relations of image blocks, she or he can forge authenticated images such that they pass tampering detection. In embedding procedure, there are two weakness that are able to cause security problems shown as follows.

\textbf{Weakness 1:} For each $b\times b$ block, $m'\cdot b^2$ bits of its $m'$ MSB layers and its corresponding $l\cdot b^2-L_a$ reference bits are fed into a hash function to generate $L_a$ authentication bits. In this operation, we find that there is no secret key participating. In other words, anyone can implement this operation.

\textbf{Weakness 2:} After generating authentication bits for each block, we need permute the $l\cdot b^2-L_a$ reference bits and $L_a$ authentication bits through a secret key and then embed $l\cdot b^2$ bits in the $l$ LSB layers of each block. Assume that the permutation key is unchanged for each block. For $l\cdot b^2$ watermark bits, the maximal permutation numbers are $(l\cdot b^2)!$. Therefore, if an attacker obtains the permutation relation of watermark bits for each block, she or he is able to forge any authenticated images. For example, for an image of mode $(6, 2)$ with $2\times 2$ block, 8 watermark bits are embedd into 2 LSB of the 4 pixels. As Fig.2 shown, the maximal permutation numbers of 8 watermark bits are $8!$. Thus, an attacker tries $8!$ times at most that he/she can obtains the permutation relation of watermark bits for each block.

\begin{figure}
\centering
\includegraphics[scale=0.5]{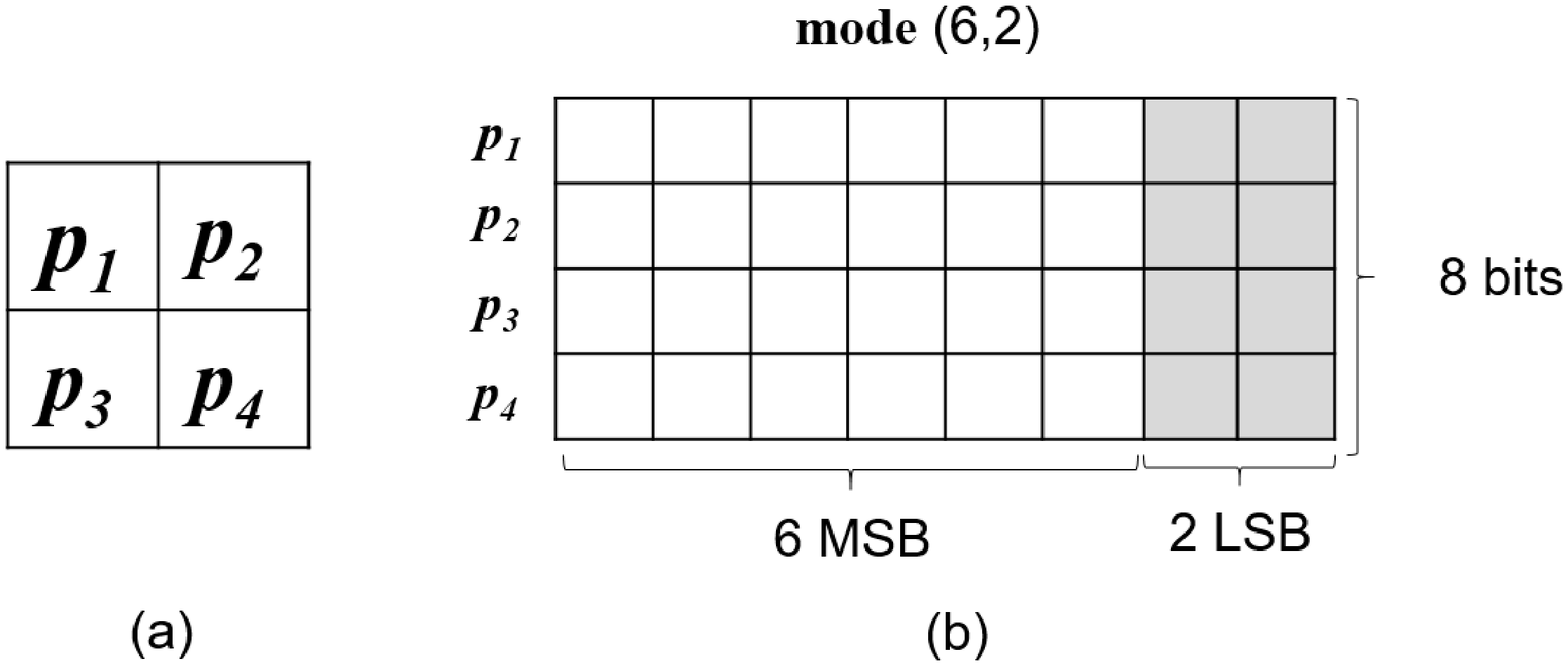}
  \caption{An image of a $2\times 2$ block (a), the layers of an block in mode $(6, 2)$ (b)}
\label{fig:Fig. 2}
\end{figure}

\textbf{Attack aim:} The aim of this attack is to obtain the equivalent permutation key (i.e., permutation relation) of $l\cdot b^2$ watermark bits in each block.

\textbf{Attack Conditions:} An attacker has two authenticated images with mode $(m, l)$.

\textbf{Step 1.} Given an authenticated image with mode $(m, l)$, we first analyze its 1st block. We extract the $l\cdot b^2$ watermark bits and separates them into two parts, $l\cdot b^2-L^*_a$ reference bits and $L^*_a$ authentication bits. The $l\cdot b^2$ bits have different permutation relations and the permutation number is expressed by the form
\begin{equation}
 (l\cdot b^2)!
\end{equation}

\textbf{Step 2.} Choose one of all permutation relations and calculate its authentication bits $L^{**}_a$. Feed $m'\cdot b^2$ bits of its $m'$ MSB layers and the $l\cdot b^2-L^*_a$ reference bits into a hash function and generate authentication bits $L_a^{**}$.

\textbf{Step 3.} Compare $L^*_a$ and $L^{**}_a$. If $L^{**}_a \neq L^*_a$, the assumed permutation is wrong, otherwise it may be right. Furthermore, we take another authenticated image to verify this permutation relation.

\textbf{Step 4.} Parallel processing all blocks. We execute the parallel process of blocks of the authenticated image from steps 2 to 3. To acquire the correct permutation of $l\cdot b^2$ watermark bits, we need test about $(l\cdot b^2)!$ times, which is not related to the image size. Table 1 lists the test number of analysis for different block sizes and embedding modes. From Table 1 we can observe that the security of Qin et al.'s scheme increases with increase of block size of the image and embedding layer of the watermark, but the security is low for small block size.

We measure the time for obtaining all the permutation relation of $l\cdot b^2$, shown in Table 1. The experiments are implemented on a personal computer with a 3.60 GHz Intel i7 processor, 8.00 GB memory, and Windows 10 operating system. Implementation software is Matlab R2015a. For the block size $2 \times 2$, the time of mode (6.2) is about $11.9953 \ s$ while the time of mode (6.3) is $186732 \ s \simeq 52 \ hours$. The time ratio of mode (6,3) to mode (6,2) is theoretically 11880, but is actually $15567$, this is because the program about mode (6,3) requires large storage space. For the block size $4 \times 4$ and mode $(6, 3)$, the maximum test number is up to $2^{202.9}$. Therefore, the larger the block size, the more test number and time the multiple stego-image attack needs to be tried.

\begin{table}[htbp]
\caption{The test number and time of multiple stego-image attack for different block sizes and embedding modes.}
\centering  
\begin{tabular}{lcccc}
\hline
Block size&Mode&Test number&Test time(s)\\ \hline
$1\times 1$&(6,2)&2&$0.0035 $ \\
$1\times 1$&(6,3)&6$\approx 2^{2.6}$&$0.0052$ \\
$2\times 2$&(6,2)&40320 $\approx 2^{15.3}$&$11.9953$ \\
$2\times 2$&(6,3)& $4.7900*10^{8} \approx 2^{28.8}$&$186732 \approx 2^{17.5}$\\
$4\times 4$&(6,2)& $2.6313*10^{35} \approx 2^{117.6}$&$/$\\
$4\times 4$&(6,3)& $1.2414*10^{61} \approx 2^{202.9}$&$/$\\ \hline
\end{tabular}
\end{table}

\section{Conclusion}
Qin et al. proposed a general image self-embedding watermark scheme for tampering recovery. But we find that, due to the block independence of watermark embedding, we can forge a new authenticated image from authenticated images via the collage attack. The simulation results verify our theoretical analysis. Furthermore, we analyze the security of Qin et al.'s scheme by using multiple stego-image attack. Because the generation of authentication bits isn't related to a secret key, once an attacker acquires the permutation relations of $l\cdot b^2$ watermark bits of all blocks, she or he can forge any authenticated images. The cost of acquiring all the permutation relations is about $(l\cdot b^2)!$, where $l$ LSB layers are used for embedding watermark, and $b \times b$ is the size of image block. Enhancing the security of fragile watermarking algorithms has been a challenge and we hope our analysis method will promote the research of fragile watermarking to some extent.


\bibliographystyle{unsrt}
\bibliography{mybibfile}   

%
%


\end{document}